\documentclass[twocolumn,prd,preprintnumbers,amsmath,amssymb,nofootinbib]{revtex4}
\usepackage{graphicx}
\usepackage{dcolumn}
\usepackage{bm}
\usepackage[dvips]{color}

\newcommand{\beq}{\begin{equation}}
\newcommand{\eeq}{\end{equation}}
\newcommand{\bq}{\begin{equation}}
\newcommand{\eq}{\end{equation}}
\newcommand{\ba}{\begin{array}}
\newcommand{\ea}{\end{array}}
\newcommand{\beqa}{\begin{eqnarray}}
\newcommand{\eeqa}{\end{eqnarray}}

\def\[{\left[}
\def\]{\right]}
\def\({\left(}
\def\){\right)}

\evensidemargin=\oddsidemargin

\def\pslash{\not{\hbox{\kern-4pt $p$}}}
\def\qslash{\not{\hbox{\kern-4pt $q$}}}
\def\lv{\not{\hbox{\kern-4pt $L$}}}
\def\lsim{\mathrel{\raise.3ex\hbox{$<$\kern-.75em\lower1ex\hbox{$\sim$}}}}
\def\gsim{\mathrel{\raise.3ex\hbox{$>$\kern-.75em\lower1ex\hbox{$\sim$}}}}
\def\ifmath#1{\relax\ifmmode #1\else $#1$\fi}
\begin{document}

 \title{A Four-site Higgsless Model with Wavefunction Mixing}

\author{ R.\ Sekhar Chivukula and Elizabeth H.\ Simmons}

 \affiliation{
 Department of Physics and Astronomy, Michigan State
       University, East Lansing, MI 48824, USA}

\date{\today}

\begin{abstract}
Motivated by models of holographic technicolor, we discuss a four-site deconstructed Higgsless model
with nontrivial wavefunction mixing. We compute the spectrum of the model, the electroweak triple gauge boson vertices, and, for
brane-localized fermions, the electroweak parameters to ${\cal O}(M^2_W/M^2_\rho)$.
We discuss the conditions under which $\alpha S$ vanishes (even for brane-localized fermions) and the  (distinct but overlapping) conditions under which the phenomenologically interesting decay
$a_1 \to W \gamma$ is non-zero and suppressed by only one power of $(M_W/M_\rho)$.
\end{abstract}

\date{\today}
 
 \maketitle


\section{Introduction}

Higgsless models of electroweak symmetry breaking \cite{Csaki:2003dt}  
may  be viewed
as ``dual" to more conventional technicolor models
\cite{Weinberg:1979bn,Susskind:1978ms} and, as such, provide a basis for constructing low-energy effective theories
to investigate the phenomenology of a strongly interacting symmetry breaking sector
\cite{He:2007ge,Hirn:2007we}. One approach to constructing such an 
effective theory, the three-site model \cite{SekharChivukula:2006cg}, includes only
the lightest of the extra vector mesons typically present in such theories -- the meson analogous to the $\rho$ in
QCD. An alternative approach is given by  ``holographic technicolor" 
\cite{Hirn:2006nt}, which potentially provides a description of the first two extra
vector  mesons -- including, in addition to the $\rho$,  the analog of the $a_1$ meson in QCD.

\begin{figure}
\begin{center}
\includegraphics[width=6cm]{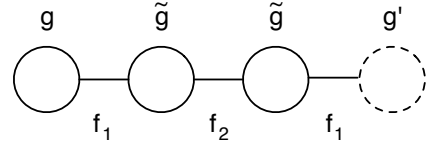}
\end{center}
\caption{The ``moose" diagram \protect\cite{Georgi:1985hf} for the $SU(2)^3 \times U(1)$ model considered in
this note. The solid circles represent $SU(2)$ groups; the dashed circle, a $U(1)$ group; the
``links",  $SU(2) \times SU(2)/SU(2)$ non-linear sigma models.
In order to
be phenomenologically realistic \protect\cite{SekharChivukula:2004mu}, we work in the limit $g,g' \ll \tilde{g}$; in this limit the model also 
has an approximate parity symmetry. We consider brane-localized fermions, which
couple only the the $SU(2) \times U(1)$ at the ends of the moose, and add an
$L_{10}$-like ``wavefunction mixing" term to mix the two strongly-coupled $SU(2)$ groups
in the middle two sites.
\label{fig:one}}
\end{figure}

In this note we consider consider a four-site ``Higgsless" model \cite{Accomando:2008jh}
illustrated, using ``moose notation"  \cite{Georgi:1985hf},  in fig. \ref{fig:one}.
We show how, once an $L_{10}$-like
``wavefunction" mixing term for the two strongly-coupled $SU(2)$ groups in the 
center of the moose is included, we can reproduce the features of 
the holographic model -- including the vanishing of the parameter $\alpha S$
for brane-localized fermions and the existence  (whether or not $\alpha S = 0$) of the potentially interesting
decay $a_1 \to W \gamma$.

\section{The Model}

The Lagrangian for the model consists of several parts. First, the usual nonlinear
sigma model link terms
\begin{align}
{\cal L}_\pi = & \frac{f^2_1}{4}\left[{\rm Tr} D^\mu \Sigma_1 D_\mu \Sigma^\dagger_1
 + {\rm Tr} D^\mu \Sigma_3 D_\mu \Sigma^\dagger_3\right] \nonumber\\
 & + \frac{f^2_2}{4}
{\rm Tr} D^\mu \Sigma_2 D_\mu \Sigma^\dagger_2~.
\label{eq:nlsm}
\end{align}
Next, the gauge-boson kinetic energies
\begin{equation}
{\cal L}_{gauge} = -\,\frac{1}{4}\left(\vec{W}^2_{0\mu\nu}
+ \vec{W}^2_{1\mu\nu}+\vec{W}^2_{2\mu\nu}+ \vec{W}^2_{3\mu \nu}\right)~,
\label{eq:gaugekinetic}
\end{equation}
where we denote the weakly-coupled $SU(2)\times U(1)$ fields by
$\vec{W}_0$ and $\vec{W}_3 \equiv B$ (by convention, $i=3$ vanishes for the charged sector), and the strongly coupled $SU(2)$ fields by $\vec{W}_{1,2}$. And finally,  there is an $L_{10}$-like mixing between the middle two sites
\begin{equation}
{\cal L}_\varepsilon = -\,\frac{\varepsilon}{2} {\rm Tr}\left[
\vec{W}_{1\mu\nu} \Sigma_2 \vec{W}^{\mu\nu}_2 \Sigma^\dagger_2\right]~,
\label{eq:l10term}
\end{equation}
where in this calculation we treat  $\varepsilon$ as an ${\cal O}(1)$ parameter. 
This model has a ``parity" (more precisely, a $G$-parity) 
symmetry in the $g=g'=0$ limit, under which
$\vec{W}^\mu_1 \to \vec{W}^\mu_2$,  $\Sigma_1 \to \Sigma_3^\dagger$, and 
$\Sigma_2 \to \Sigma_2^\dagger$. In the limit $f_2 \to \infty$,\footnote{For fixed
values of $2/f^2_1 + 1/f^2_2$, see eqn. (\protect\ref{eq:GF}).} this model
reduces to the three-site model considered in \cite{SekharChivukula:2006cg}.

In unitary gauge (with  $\Sigma_1=\Sigma_2=\Sigma_3 \equiv {\cal I}$), the
${\cal L}_\varepsilon$ term above corresponds to wavefunction-mixing of 
the fields $\vec{W}_i$, 
%
\begin{equation}
{\cal L} = -\,\frac{1}{4} \vec{W}_{i\mu\nu} \tilde{Z}_{ij} \vec{W}_j^{\mu\nu} 
-\frac{1}{2} \vec{W}_{i\mu} M^2_{ij} \vec{W}^\mu_j~,
\label{eq:nsite}
\end{equation}
with 
\begin{equation}
\tilde{Z} = \begin{pmatrix}
1 &  &  &   \\
  & 1 &  \varepsilon &   \\
 & \varepsilon & 1 &  \\
 &   &   & 1
\end{pmatrix}~.
\end{equation}
To avoid ghosts, we require $\tilde{Z}$ to 
be positive-definite, and hence $|\varepsilon| < 1$.

\section{Masses and Mixing Angles}

The eigenstates corresponding to the quadratic part of Lagrangian in eqn. (\ref{eq:nsite})
satisfy the generalized eigenvalue equation
\begin{equation}
M^2 \vec{v}_n = m^2_n \tilde{Z} \vec{v}_n~,
\label{eq:eigeneqn}
\end{equation}
where $\vec{v}_n$ is a vector in site-space with components $v_n^i$.  The superscript 
$i$ labels the sites, running from 0 to 2 for charged-bosons ($n = W^\pm, \rho^\pm, a_1^\pm$), and 0 to 3 for neutral
ones ($n = Z^0, \rho^0, a_1^0, \gamma$).
If we choose eigenvectors normalized by $\vec{v}^T_n \tilde{Z} \vec{v}_m = \delta_{nm}$,
the gauge-eigenstate ($W_\mu^i$) and mass-eigenstate ($W'_{n\mu}$)
fields are related by
\begin{equation}
W^i_\mu = \sum_n v^i_n W_{n\mu}' ~.
\label{eq:gauge-states}
\end{equation}

\subsection{The $g=g'=0$ Limit}

Consider first  the $g=g'=0$ limit, in which we can determine the leading
contributions to the heavy gauge-boson masses. Due to the parity symmetry in this
limit, we expect the eigenvectors to be proportional to $\vec{W}^\mu_1 \pm \vec{W}^\mu_2$.
Applying the normalization condition $\vec{v}^T_n \tilde{Z} \vec{v}_m = \delta_{nm}$, we find a parity-even
eigenvector (the ``$\rho$")
\begin{equation}
\vec{\rho}^\mu = \frac{1}{\sqrt{2(1+\varepsilon)}}\left(
\vec{W}_1^\mu + \vec{W}_2^\mu \right)~,
\label{eq:rhovector}
\end{equation}
with mass
\begin{equation}
m^2_\rho = \frac{\tilde{g}^2}{4} \frac{f^2_1}{1+\varepsilon}~,
\label{eq:rhomass}
\end{equation}
and a parity-odd eigenvector (the ``$a_1$")
\begin{equation}
\vec{a}_1^\mu = \frac{1}{\sqrt{2(1-\varepsilon)}} \left(
\vec{W}_1^\mu - \vec{W}_2^\mu \right)~,
\label{eq:a_1vector}
\end{equation}
with mass
\begin{equation}
m^2_{a_1} = \frac{\tilde{g}^2}{4} \frac{f^2_1 + 2 f^2_2}{1-\varepsilon}~.
\label{eq:a_1mass}
\end{equation}
We note that the $\rho$ and $a_1$ are degenerate for
\begin{equation}
\varepsilon = -\, \frac{f^2_2}{f^2_1 + f^2_2}~,
\label{eq:degeneracy}
\end{equation}
a value satisfying
the constraint $|\varepsilon| < 1$. As $\varepsilon$ becomes
more negative, the $a_1$ becomes lighter than the $\rho$.

\subsection{The Photon}

Examining the eigenvalue eqn. (\ref{eq:eigeneqn}) we see that the wavefunction factor
$\tilde{Z}$ affects the normalization of a massless eigenvector, but not the orientation. We see,
therefore, that the photon must be of the form
\begin{equation}
A_\mu = \frac{e}{g} W^3_{0\mu} + \frac{e}{\tilde{g}} W^3_{1\mu} +
\frac{e}{\tilde{g}} W^3_{2\mu} + \frac{e}{g'} B_\mu~,
\label{eq:photonvector}
\end{equation}
or
\begin{equation}
(v_\gamma)^T = \left(\frac{e}{g}\ ,\ \ \frac{e}{\tilde{g}} \ , \ \  \frac{e}{\tilde{g}} \ , \ \ \frac{e}{g'}\right)~.
\label{eq:photon}
\end{equation}
The electric charge $e$ is, then, determined from the normalization condition to be 
\begin{equation}
\frac{1}{e^2} = \frac{1}{g^2} + \frac{1}{{g'}^2} + \frac{2(1+\varepsilon)}{\tilde{g}^2}~.
\label{eq:charge}
\end{equation}
Examining the photon-couplings, we see that the unbroken gauge-generator
has the expected form $Q= T^3+T^3_1 + T^3_2 + Y$.

\subsection{The $W$-boson}

Next, we consider a perturbative evaluation of the electroweak boson eigenvectors and eigenvalues, 
computed in powers of $x = g/\tilde{g}$. We start with the $W$-boson; the
charged-boson mass matrix is given by
\begin{equation}
M^2_W = \frac{\tilde{g}^2}{4}
\begin{pmatrix}
x^2 f^2_1 & -x f^2_1 & 0 \\
-x f^2_1 & f^2_1 + f^2_2 & - f^2_2 \\
0 & -f^2_2 & f^2_1+f^2_2
\end{pmatrix}~.
\label{eq:mwsq}
\end{equation}
To ${\cal O}(x^2)$ we find
\begin{align}
v^0_W & = \left[1-\frac{f^4_1 + 2(1+\varepsilon) f^2_1 f^2_2 + 2(1+\varepsilon) f^4_2}{2(f^2_1+2 f^2_2)^2}\,x^2~,
\right] \nonumber \\
v^1_W & =  x\,\frac{f^2_1+f^2_2}{f^1_1+2 f^2_2}W_1~, \\
 v^2_W &= x\,\frac{f^2_2}{f^2_1+2 f^2_2} W_2~,\nonumber
\end{align}
where we have computed, but do not display, the corrections of ${\cal O}(x^3)$ to the
last two components.
For the corresponding eigenvalue we find
\begin{equation}
m^2_W = \frac{g^2}{4} \frac{f^2_1 f^2_2}{f^2_1+2f^2_2}
\left[1-\frac{f^4_1 + 2(1+\varepsilon) f^2_1 f^2_2 + 2(1+\varepsilon) f^4_2}{(f^2_1+2 f^2_2)^2}\,x^2
\right]~.
\end{equation}

\subsection{The $Z$-boson}

The neutral gauge-boson mass matrix is
\begin{equation}
M^2_Z = 
\begin{pmatrix}
x^2 f^2_1 & -x f^2_1 & 0 & 0\\
-x f^2_1 & f^2_1 + f^2_2 & - f^2_2 & 0 \\
0 & -f^2_2 & f^2_1+f^2_2 & -x\tan\theta f^2_1 \\
0 & 0 & -x \tan\theta  f^2_1 & x^2 \tan^2\theta f^2_1
\end{pmatrix}~.
\label{eq:mzsq}
\end{equation}
where we have defined the angle $\theta$ by
$g'/g \equiv \tan \theta$.
Note that $\theta$ is the {\it leading order} weak mixing angle; we will later define a weak mixing angle $\theta_Z$ that is better suited to comparison with experiment.
We have computed the $Z$-boson eigenvector to ${\cal O}(x^3)$ -- as the 
result is complicated, and the algebra unilluminating, we do not reproduce it here. For the $Z$-boson 
mass, we find
\begin{widetext}
\begin{equation}
m^2_Z = \frac{g^2}{4\cos^2\theta} \frac{f^2_1 f^2_2}{f^2_1+2f^2_2} 
\left[
1-\frac{(3-\varepsilon) f^4_1+4(1+\varepsilon)(f^2_1 f^2_2 + f^4_2) +
(1+\varepsilon)(f^2_1+2 f^2_2)^2 \cos4\theta}{4(f^2_1+2f^2_2)^2}\,x^2 \sec^2\theta
\right]~.
\label{eq:m2z}
\end{equation}
\end{widetext}

\section{The Electroweak Parameters}

From eqn. (\ref{eq:gauge-states}), we can compute the couplings of the mass-eigenstate
electroweak gauge-bosons to fermions. For
brane-localized  fermion couplings of the form
\begin{equation}
{\cal L}_f = g_0 \vec{J}^\mu_L \cdot  \vec{W}^0_\mu
+ g' J^\mu_Y B_\mu~,
\end{equation}
 we find  the mass-eigenstate $W$-boson couplings
$g_W^f = g_0 v^0_W$ and the $Z$-boson couplings
\begin{equation} 
g^f_Z = g v^0_Z I_3 + g' v^3_Z Y = g I_3(v^0_Z - \tan\theta v^3_Z) + g' v^3_Z {\cal Q}~.
\label{eq:zcouplings}
\end{equation}
We may then compute the on-shell precision electroweak parameters at tree-level to
${\cal O}(x^2)$, using the definitions and procedures outlined in \cite{Chivukula:2004af,SekharChivukula:2004mu}. The values of electric charge, eqn. (\ref{eq:charge}),
and $m^2_Z$, eqn. (\ref{eq:m2z}),  are given above, and we find the Fermi constant
\begin{equation}
\sqrt{2} G_F = \frac{1}{v^2} = \frac{2}{f^2_1} + \frac{1}{f^2_2}~,
\label{eq:GF}
\end{equation}
where $v\approx 246$ GeV. 
  
The only non-zero precision electroweak parameter parameter is
  $\alpha S$ \cite{Peskin:1991sw},  for which we find
\begin{equation}
\frac{\alpha S}{4 s^2} = \frac{\varepsilon f^4_1 + 2(1+\varepsilon) f^2_1 f^2_2 + 2 f^4_2 (1+\varepsilon)}
{(f^2_1 + 2f^2_2)^2}\, x^2~,
\end{equation}
As expected \cite{Hirn:2006nt,Hirn:2007we}, we can choose $\varepsilon$ so that $\alpha S$ vanishes for any given value of  $f_1/f_2$
\begin{equation}
\varepsilon \to -\, \frac{2(f^4_2 + f^2_1 f^2_2)}{f^4_1 + 2 f^2_1 f^2_2 + 2 f^4_2}~,
\label{eq:zeroS}
\end{equation}
while satisfying $\vert\varepsilon\vert < 1$. 

Note, however, that the value of 
the low-energy parameter  $\vert\varepsilon\vert$ that makes $\alpha S$  vanish
is of order one, larger than would be expected by naive dimensional
analysis \cite{Georgi:1992dw}.  This result is consistent with investigations
of continuum 5d effective theories \cite{Hong:2006si,Agashe:2007mc}, and
with  investigations of plausible conformal technicolor  ``high-energy completions" 
of this model using Bethe-Salpeter methods \cite{Harada:2005ru,Kurachi:2006mu}, 
both of  which  suggest that $\alpha S >0$ 
and that it may not be possible 
to achieve very small values of $\alpha S$. We note also that the result is consistent with  
the expectation of \cite{Appelquist:1998xf,Appelquist:1999dq}, since
the value of $\varepsilon$ required for $\alpha S$ to vanish results in
axial-vector mesons which are lighter than the vector mesons.\footnote{An alternative
approach, Degenerate BESS \protect\cite{Casalbuoni:1995yb,Casalbuoni:1995qt}, produces
degenerate vector and axial mesons and $\alpha S=0$ using a different theory without unitarity delay 
\cite{SekharChivukula:2004mu}
-- see ``case I" described in \protect\cite{Chivukula:2003wj}. }  

\section{Triple Boson Vertices}

\subsection{Electroweak Vertices}

Consider the electroweak vertices $\gamma WW$ and $ZWW$.
To leading order, in the absence of CP-violation, the triple gauge boson vertices
may be written \cite{Hagiwara:1986vm}
\begin{eqnarray}
{\cal L}_{TGV} & = & - ie\frac{c_Z }{s_Z}\left[1+\Delta\kappa_Z\right] W^+_\mu W^-_\nu Z^{\mu\nu} \nonumber \\
& -&  ie \left[1+\Delta \kappa_\gamma\right] W^+_\mu W^-_\nu A^{\mu\nu} \nonumber \\
&-& i e \frac{c_Z}{s_Z} \left[ 1+\Delta g^Z_1\right](W^{+\mu\nu}W^-_\mu - W^{-\mu\nu} W^+_\mu)Z_\nu \nonumber \\
& -&  ie (W^{+\mu\nu}W^-_\mu - W^{-\mu\nu} W^+_\mu) A_\nu~, 
\label{eq:3point}
\end{eqnarray}
where the two-index tensors denote the Lorentz field-strength
tensor of the corresponding field. In the standard model, 
$\Delta\kappa_Z = \Delta\kappa_\gamma = \Delta g^Z_1 \equiv 0$.
Note that the expressions for $\kappa_Z$ and $g^Z_1$ involve
$c_Z \equiv \cos \theta_Z$ and $s_Z \equiv \sin \theta_Z$, as defined by
\begin{equation}
c^2_Z s^2_Z = \frac{e^2}{4 \sqrt{2} G_F M_Z^2},
\end{equation}
rather than the leading order mixing angle $\theta$.

Let us begin with the coupling of the photon of the form
$(W^{+\mu\nu}W^-_\mu - W^{-\mu\nu} W^+_\mu) A_\nu$.  In terms of the wavefunctions
$v_{\gamma, W}$, this coupling is proportional to
\begin{equation}
g_\gamma = \sum_{i,j}  g_i v^i_\gamma v^i_W \tilde{Z}_{ij} v_W^j~.
\label{eq:ggamma}
\end{equation}
From eqn. (\ref{eq:photon}), we have $g_i v^i_\gamma \equiv
e$ and therefore, by applying the normalization condition $\vec{v}^T_W \tilde{Z} \vec{v}_W = 1$, we obtain
$g_\gamma \equiv e$ independent of any choice of the four-site parameters
 --- as required by gauge-invariance and consistent with the form
 of eqn. (\ref{eq:3point}).

Next, we evaluate $\Delta \kappa_\gamma$, with
\begin{equation}
e\,[1+\Delta\kappa_\gamma]   =  \sum_{i,j} g_i (v^i_W)^2 \tilde{Z}_{ij} v^j_\gamma = e\, \sum_{i,j} \frac{g_i}{g_j} (v^i_W)^2 \tilde{Z}_{ij}~,
\end{equation}
for which we calculate
\begin{equation}
\Delta \kappa_\gamma = \frac{\varepsilon\, f^4_1}{(f^2_1+2 f^2_2)^2}\, x^2
= \frac{\varepsilon\, v^4}{f^4_2}\, x^2~.
\end{equation}
Note that this vanishes in the absence of wavefunction mixing ($\varepsilon \to 0$), and also in the
``three-site" limit ($v/f_2 \to 0$), as consistent with  \cite{SekharChivukula:2006cg}.

Similarly we may compute $\Delta g^Z_1$ and $\Delta \kappa_Z$,
and we find
%
\begin{eqnarray}
&\Delta g^Z_1&  =  \Delta \kappa_Z +  \frac{\varepsilon f^4_1 \tan^2\theta_Z\, x^2}{(f^2_1+2f^2_2)^2}~,\\
&=& - \, \frac{( \varepsilon s^2_Z f^4_1 + (1+\varepsilon)f^2_1 f^2_2+(1+\varepsilon)f^4_2) }{(f^2_1+2 f^2_2)^2 \cos(2\theta_Z)} \frac{x^2}{c^2_Z} ~,\nonumber
\end{eqnarray}
%
where the difference between $\theta$ and $\theta_Z$ is irrelevant to this order. Note that $\Delta g^Z_1-\Delta \kappa_Z$
vanishes when $\varepsilon \to 0$, and also, as expected  \cite{SekharChivukula:2006cg}, in the ``three-site" limit $f_2 \to \infty$.

\subsection{$\rho,\,a_1 \to W + \gamma$}

Finally, we consider the $(\rho,a_1) - W - \gamma$ couplings that motivated this
study. Electromagnetic gauge-invariance implies
that the coupling of the form $(\rho^{+\mu\nu}W^-_\mu - W^{-\mu\nu} \rho^+_\mu) A_\nu$
must vanish. Analogous to eqn. (\ref{eq:ggamma}) we find that  the $\rho - W - \gamma$ 
and $a_1 - W - \gamma$ couplings
of this form are proportional to $\vec{v}^T_W \tilde{Z} \vec{v}_{\rho,a_1} \equiv 0$, and
therefore vanish identically.

There is no reason, however, that  terms proportional
to $(\rho^+_\mu,\, a^+_{1\mu})\,  W^-_\nu A^{\mu\nu}$ must vanish \cite{Hirn:2006nt,Hirn:2007we}. In this case, we find
\begin{equation}
e\, \kappa_{\gamma W \rho} = \sum_{i,j} g_i v^i_W v^i_\rho  \tilde{Z}_{ij} v^j_\gamma = e \sum_{i,j} \frac{g_i}{g_j} v^i_W v^i_\rho  \tilde{Z}_{ij}~,
\end{equation}
and similarly for the $a_1$.
Computing these couplings to ${\cal O}(x^3)$, 
we find
\begin{eqnarray}
\kappa_{\gamma W \rho} & = & -\, \frac{\varepsilon(1+\varepsilon)^{3/2} f^4_1}
{2 \sqrt{2}(f^2_1 + 2 f^2_2) 
(\varepsilon f^2_1 + (1+\varepsilon) f^2_2)}\, x^3~,
\label{eq:rhoWgamma}\\
\kappa_{\gamma W a_1}&  = & \frac{\sqrt{2}\, \varepsilon\, v^2}{\sqrt{1-\varepsilon}\, f^2_2}\, x~.
\label{eq:a1Wgamma}
\end{eqnarray}
Note that both couplings vanishes in the $\varepsilon \to 0$ and $f_2 \to \infty$ limits.
Furthermore, while the $\rho - W-\gamma$ coupling is typically small
(${\cal O}(x^3)$), we find the $a_1 - W - \gamma$ coupling is only
suppressed by $x$, consistent with 
\cite{Hirn:2006nt,Hirn:2007we}.  If the value of $\varepsilon$ corresponds (\ref{eq:zeroS}) to $\alpha S = 0$, then $\kappa_{\gamma W a_1}$ is
\begin{equation}
\kappa_{\gamma W a_1} = - \frac{2 \sqrt{2} v^2 (f_1^2 + f_2^2)\ x}{(f_1^2 + 2 f_2^2)\sqrt{f_1^2 + 2 f_1^2 f_2^2 + 2 f_2^2}} .
\end{equation}
As mentioned earlier, for this value of $\varepsilon$, the $a_1$ state is lighter than the $\rho$.

\section{Summary}

We have introduced a deconstructed Higgsless model with four sites and non-trivial wavefunction mixing, and have shown that it exhibits key features of holographic technicolor \cite{Hirn:2007we, Hirn:2006nt}.  The electroweak parameter $\alpha S$ vanishes for a value of the wavefunction mixing at which the $a_1$ is lighter than the $\rho$ -- even if all fermions are brane-localized.  Furthermore, the model includes  the decay $a_1 \to W \gamma$, suppressed by only one power of $(M_W / M_{\rho})$, in contrast with an $(M_W/M_\rho)^3$ suppression of the decay $\rho \to W \gamma$.  These decays are of potential phenomenological interest at LHC.

\section{Acknowledgements}
This work was supported in part by the US National Science Foundation under
grant  PHY-0354226. The authors acknowledge the hospitality and support of the
    the Aspen Center for Physics, where this manuscript was completed.
We thank Tom Appelquist, Veronica Sanz, Johannes Hirn, and Adam Martin
for useful discussions.

\end{document}